\documentclass[12pt,a4paper]{article}
\usepackage{amsfonts,latexsym}
\usepackage{amsmath,amssymb}
\usepackage{graphicx,color}

\renewcommand{\thefootnote}{\fnsymbol{footnote}}

\begin{document}

\vspace{12mm}

\begin{center}
{{{\Large {\bf Scale-invariant scalar spectrum from the nonminimal derivative coupling with fourth-order  term}}}}\\[10mm]

{Yun Soo Myung\footnote{e-mail address: ysmyung@inje.ac.kr} and Taeyoon Moon\footnote{e-mail address: tymoon@inje.ac.kr}}\\[8mm]

{Institute of Basic Sciences and Department  of Computer Simulation, Inje University Gimhae 621-749, Korea\\[0pt]}

\end{center}
\vspace{2mm}

\begin{abstract}
An exactly scale-invariant spectrum of scalar perturbation generated
during de Sitter spacetime is found from the gravity model of the
nonminimal derivative coupling with fourth-order term. The
nonminimal derivative coupling term generates a healthy (ghost-free)
fourth-order derivative term, while the fourth-order term provides
an unhealthy (ghost) fourth-order derivative term. The
Harrison-Zel'dovich
 spectrum obtained from Fourier transforming the fourth-order propagator in
de Sitter space is recovered   by computing the power spectrum in
its momentum space directly. It shows  that this model provides a
truly scale-invariant spectrum, in addition to the Lee-Wick scalar
theory.

\end{abstract}
\vspace{5mm}


\vspace{1.5cm}

\hspace{11.5cm}{Typeset Using \LaTeX}
\newpage
\renewcommand{\thefootnote}{\arabic{footnote}}
\setcounter{footnote}{0}

\section{Introduction}

In the evolutionary phase  of density inhomogeneities, one often
makes the simplifying assumption that the primordial power spectrum
has the simple power-law expression like ${\cal
P}(k)=A_s(k/k_*)^{n_s-1+\frac{\alpha_s}{2}\ln(k/k_*)}$ with $k_*$ a
pivot scale. The case of $n_s=1 $ and $ \alpha_s=0$ corresponds to
the Harrison-Zeld'ovich (HZ)
spectrum~\cite{Pandolfi:2010dz,Tsujikawa:2012mk,Creminelli:2010ba}
and this has been ruled out by different
datasets~\cite{Lanusse:2014sra}. In the picture of slow-roll
inflation (quasi-de Sitter expansion), the momentum dependence at
any given time arises as a consequence of the time dependence of $H$
and $\dot{\phi}$ compared to the de Sitter (dS) expansion of
constant $H$ and $\dot{\phi}=0$.  Recent data from Planck has shown
that the scalar spectrum is a nearly scale-invariant one with the
amplitude  $A_s=\frac{1}{2\epsilon_* M^2_{\rm
P}}(H_*^2/2\pi)^2=(2.441\pm 0.092)\times 10^{-9} $, implying that it
is approximately 1(more precisely, $n_s=0.9603\pm
0.0073$)~\cite{Ade:2013zuv}.

On the other hand, it is worth noting that the power spectrum of a
massless minimally coupled (mmc) scalar  in dS expansion takes  the
form of $(H/2\pi)^2[1+(k/aH)^2]$. It reduces to the HZ
scale-invariant spectrum of $(H/2\pi)^2$ in the superhorizon region
of $k \ll aH$, whereas it leads to  $(k/2\pi a)^2$ in the subhorizon
region of $k\gg aH$~\cite{Baumann:2009ds}. We note here that the
latter is just the spectrum of a massless conformally coupled (mcc)
scalar generated during dS expansion.  This happens to the tensor
spectrum too.   Clearly, the HZ scale-invariant  spectrum is related
to the two-point function which is logarithmically growing for
largely separated points in dS space. Recently, it was shown that
the IR growing of the two-point function is physical because Fourier
transforming logarithmic two-point function can lead to the HZ
spectrum when one tames logarithmic divergence by using
Ces\`{a}ro-summability technique~\cite{Youssef:2012cx}. This implies
that one could obtain the HZ spectrum through  an IR regularization
procedure.

More recently, the authors have obtained  the HZ spectrum
$(H/2\pi)^2$ from the Lee-Wick model of a fourth-order derivative
scalar theory in dS spacetime~\cite{Myung:2014mla}. Here we have
obtained the fourth-order propagator as the inverse of the Lee-Wick
(LW) operator $\Delta_{\rm
LW}=-\frac{1}{M^2}(\bar{\nabla}^4-M^2\bar{\nabla}^2)$ with $M^2$
mass parameter. The operator of ``$-M^2\Delta_{\rm LW}$'' becomes
the Weyl operator $\Delta_4=\bar{\nabla}^4-2H^2\bar{\nabla}^2$ in dS
spacetime~\cite{Mazur:2001aa} which is a conformally covariant
operator only when choosing $M^2=2H^2$.  The HZ
 spectrum  obtained by Fourier transforming the  propagator  in
dS spacetime was confirmed  by computing the power spectrum directly
in momentum space.  Also, the scale-invariant tensor spectrum
generated during dS inflation could be found from the conformal
gravity of $\sqrt{-g}C^2$~\cite{Myung:2015vya}. This suggests
strongly that a fourth-order derivative theory related to conformal
symmetry provides the HZ scale-invariant spectrum in whole dS
spacetime.

In this work, we wish to look for another model which may  provide a
HZ scale-invariant spectrum. This would be  the nonminimal
derivative
coupling~\cite{Amendola:1993uh,Capozziello:1999uwa,Saridakis:2010mf,Germani:2010gm,Germani:2011ua}
with fourth-order  term because this coupling provides $-3\xi
H^2\bar{\nabla}^2$-term naturally for dS inflation. If a priori
arbitrary parameter $\xi$ is tuned to be 2/3, one could obtain the
scale-invariant spectrum. In this case, this model is more
attractive than the LW model. Although the computation of the
slow-roll inflation (quasi-dS expansion) is promising to compare
with observation data, we here compute power spectrum generated
during dS inflation because the computation is more easy and
intuitive than those in quasi-dS expansion.

\section{Gravity model}

We start with   the  gravity model  whose action is given by
\begin{eqnarray} \label{enf}
S_{\rm ENF}=S_{\rm E}+S_{\rm NF}=\int d^4x
\sqrt{-g}\Big[\Big(\frac{R}{2\kappa}-\Lambda\Big)+\frac{1}{2M^2}\Big(\xi
G_{\mu\nu}\partial^\mu\phi\partial^\nu\phi-(\nabla^2\phi)^2\Big)\Big],
\end{eqnarray}
where $\kappa=8\pi G=1/M^2_{\rm P}$, $M_{\rm P}$ being the reduced
Planck mass, and  $M^2$ is a coupling  parameter with dimension of
mass squared. Here $G_{\mu\nu}=R_{\mu\nu}-Rg_{\mu\nu}/2$ is the
Einstein tensor  and $\xi$ is a coefficient to be adjusted as
$\frac{2}{3}$. $\Lambda$ is introduced as a positive cosmological
constant to obtain dS spacetime, instead of potential $V$. We note
that  the scalar $\phi$ has dimension of mass. The former term in
the last parenthesis is the nonminimal derivative
 coupling (NDC) term ~\cite{Amendola:1993uh,Capozziello:1999uwa,Saridakis:2010mf,Germani:2010gm,Germani:2011ua} and the latter denotes the
fourth-order  kinetic (FK) term. We note that  the former generates
no more degrees of freedom (DOF)  than general relativity
canonically coupled to a scalar field, while the latter generates a
new scalar DOF because it is a fourth-order derivative term. It may
be noted  that the former generates  healthy (ghost-free) higher
derivative terms, while the latter is unhealthy (ghost) higher
derivative terms. At the first sight, the combined action $S_{\rm
NF}$ seems to be an unbalanced scalar model.
 For
reference, we introduce  the LW  model  given
by~\cite{Lee:1969fy,Lee:1970iw,Cai:2008qw}
\begin{equation}\label{lw}
S_{\rm LW}=-\frac{1}{2}\int
d^4x\sqrt{-g}\Big[g_{\mu\nu}\partial^\mu\phi\partial^\nu\phi+\frac{1}{M^2}(\nabla^2\phi)^2\Big],
\end{equation}
where the first term denotes the canonically coupled (CC) term.

Now we wish to explain the cosmological relevance of our model
$S_{\rm NF}$.
 The
NDC was introduced firstly  by noting  that the friction is enhanced
gravitationally at higher energies in the slow-roll
inflation~\cite{Germani:2010gm}. Actually, the NDC  makes any
potential adequate for inflation without introducing dangerous
higher time derivative (ghost state)~\cite{Germani:2011ua}.  The NDC
flattens a steep (non-flat) potential effectively as well as it
increases friction. Also, we note that this kind of coupling could
be view as either  a subgroup of Horndeski
theory~\cite{Horndeski:1974wa} or the most general scalar-tensor
theory which is regarded as a ghost-free
theory~\cite{Deffayet:2009mn}. These theories were used to construct
theoretically consistent model of
inflation~\cite{Tsujikawa:2014mba}. Here, we wish to explore the
role of NDC in the dS inflation by reminding that $G_{\mu\nu}$ takes
the simple form of $-3H^2g_{\mu\nu}$ during dS phase.

At this stage, we would like to mention the inclusion of  FK term.
In this case, one immediately  ask the question of ``can the model
(\ref{enf}) be a consistent theory?"  because the FK term generates
the ghost in Minkowski spacetime.  The existence of the ghost may
imply the inconsistency of the model. For simplicity, we consider
the LW theory (\ref{lw})  because it includes the FK term too. It is
worth noting that this model was employed successfully to derive a
bouncing universe to avoid singularity  and a scale-invariant
spectrum~\cite{Cai:2008qw}. This was possible to occur because
(\ref{lw}) can be rewritten by two fields: one is a normal scalar
 $\hat{\phi}$ and the other is a LW (ghost) scalar $\tilde{\phi}$.  The LW perturbation theory is
a higher derivative theory and thus, it contains propagators with
wrong sign residues about the new poles. Lee and Wick have  provided
a prescription for handling this issue~\cite{Lee:1969fy,Lee:1970iw}.
The LW-particles associated with these new poles are not in the
spectrum, but  $\tilde{\phi}$ decays to ordinary degrees of freedom
instead. Their resummed propagators do not satisfy the usual
analyticity properties since the poles are on the physical sheet.
They  have proposed deforming integration contours in the Feynman
diagrams so that there is no catastrophic exponential growth as time
increases. This amounts to a future boundary condition and thus, the
LW theory violates the usual causal conditions. While the Lee-Wick
interpretation is peculiar, it seems to be consistent, at least in
perturbation theory, and predictions for physical observables can be
made order by order in perturbation theory.

We remind the reader that for $\xi=2/3$, the contribution from the
ghost (mcc scalar)  is found to be essential to obtain a
scale-invariant spectrum by  canceling  of time-dependent term
arising from the healthy (mmc) scalar in dS spacetime. Even though
the FK term causes the ghost state, it is necessary to have a
scale-invariant scalar spectrum in the dS inflation because it
provides a scalar degrees of freedom other than $\phi$. In the case
of $\xi\not=2/3$, $S_{\rm NF}$ implies two scalars: one is a healthy
massless scalar and the other is an unhealthy massive scalar. In
this case, there is no cancellation between two scalars and thus,
the scalar power spectrum is not positive definite (unitary).

 Varying the action (\ref{enf}) with respect to  the
metric tensor $g_{\mu\nu}$ leads to  the Einstein equation
\begin{equation} \label{einseq}
G_{\mu\nu}+\kappa \Lambda g_{\mu\nu} =\kappa T^{\rm
NF}_{\mu\nu}=\kappa (T^{\rm NDC}_{\mu\nu}+T^{\rm FK}_{\mu\nu}),
\end{equation}
where two energy-momentum tensors are given by
\begin{eqnarray}
M^2T_{\mu\nu}^{\rm
NDC}&=&\xi\Big[\frac{1}{2}R\nabla_{\mu}\phi\nabla_{\nu}\phi
-2R_{(\nu}~^\rho\nabla_{|\rho|}\phi\nabla_{\mu)}\phi
+\frac{1}{2}G_{\mu\nu}(\nabla\phi)^2-R_{\mu\rho\nu\sigma}\nabla^{\rho}\phi\nabla^{\sigma}\phi
\nonumber\\&-&\nabla_{\mu}\nabla^{\rho}\phi\nabla_{\nu}\nabla_{\rho}\phi
+(\nabla_{\mu}\nabla_{\nu}\phi)\nabla^2\phi\nonumber\\
\label{em1}&+&g_{\mu\nu}\Big(R^{\rho\sigma}\nabla_{\rho}\phi\nabla_{\sigma}\phi-\frac{1}{2}(\nabla^2\phi)^2
+\frac{1}{2}(\nabla^{\rho}\nabla^{\sigma}\phi)\nabla_{\rho}\nabla_{\sigma}\phi
\Big)\Big] \end{eqnarray}
 and
\begin{eqnarray}
M^2T^{\rm
FK}_{\mu\nu}&=&-\nabla_{\mu}(\nabla^2\phi)\nabla_{\nu}\phi-\nabla_{\nu}(\nabla^2\phi)\nabla_{\mu}\phi
+g_{\mu\nu}\Big[\nabla_{\rho}(\nabla^2\phi)\nabla^{\rho}\phi+\frac{1}{2}(\nabla^2\phi)^2\Big].
\end{eqnarray}
We observe that $T^{\rm NDC} \sim T^{\rm FK}$ when one disregards
curvature coupled derivative terms in (\ref{em1}).

On the other hand, the scalar equation for the action (\ref{enf}) is
given by
\begin{equation} \label{s-eq}
\nabla^4\phi+\xi G^{\mu\nu}\nabla_\mu\nabla_\nu \phi=0.
\end{equation}

In order to see how the universe evolves with (\ref{enf}), we
consider the spatially flat FRW spacetime by choosing cosmic time
$t$ as
\begin{equation}
ds^2_{\rm FRW}=-dt^2+a(t)^2(dx^2+dy^2+dz^2).
\end{equation}
Replacing $\Lambda$ by the potential  $V$, two Friedmann equations
take complicated forms
\begin{eqnarray}
H^2&=&\frac{\kappa}{3}\Big[\rho_{\rm NDC}+V+\rho_{\rm FK}\Big] \nonumber \\
&=&\frac{\kappa}{3}\Big[\frac{9}{2}\frac{H^2}{M^2}\dot{\phi}^2+V+\frac{1}{M^2}\Big(3\dot{H}\dot{\phi}^2+\dddot{\phi}\dot{\phi}-\frac{9}{2}H^2\dot{\phi}^2-\frac{1}{2}\ddot{\phi}^2\Big)\Big],\label{Heq}\\
\dot{H}
&=&-\frac{\kappa}{2}\Big[\Big(\frac{3H^2}{M^2}-\frac{\dot{H}}{M^2}\Big)\dot{\phi}^2+\frac{1}{M^2}\Big(6\dot{H}\dot{\phi}^2+2\dddot{\phi}\dot{\phi}+4H\ddot{\phi}\dot{\phi}\Big)\Big],\label{dHeq}
\end{eqnarray}
where the overdot ($\dot{{}}$) denotes the differentiation with
respect to time $t$.  We observe from (\ref{Heq})
 that even though fourth-order derivative terms are generated
in $T^{\rm NDC}_{\mu\nu}$, there is no ghost state in $\rho_{\rm
NDC}$ which means that any dangerous higher time-derivative is not
generated. This is why one favors the NDC term to enhance friction
effects in the slow-roll
inflation~\cite{Germani:2010gm,Germani:2011ua}.  On the contrary,
$T^{\rm FK}_{\mu\nu}$ generates ghost states in $\rho_{\rm FK}$
because of fourth-order derivative term. Hence, the inclusion of
$\rho_{\rm FK}$ is problematic to obtain a solution of the universe
evolution. This is a basic difference between NDC and FK in the FRW
universe.

The scalar equation is given by
\begin{eqnarray}
\ddddot{\phi}+6H\dddot{\phi}&+&(12H^2+6\dot{H})\ddot{\phi} \nonumber  \\
&+&(9H^3+15H\dot{H}+3\ddot{H})\dot{\phi}+M^2\frac{dV}{d\phi}=0.\label{seq}
\end{eqnarray}
Requiring  the slow-roll conditions of $\dot{\phi}^2<2H^2$ and
$\ddot{\phi}<3H\dot{\phi}$ with neglecting all higher-order time
derivative terms from FK term, Eqs. (\ref{Heq}) and (\ref{seq}) may
be written approximately as
\begin{eqnarray}
&&H^2\simeq\frac{\kappa}{3}V,\\
&&3H\dot{\phi}\simeq-\frac{M^2}{3H^2}\frac{dV}{d\phi},
\end{eqnarray}
which seem  to be oversimplified equations. This means that  it is
very difficult to find the corresponding slow-roll equations if one
includes the FK term.  However, one always  obtains the dS solution
for $\phi$=const and $V=\Lambda$ because two Friedmann and scalar
equations reduce to $H^2=\kappa \Lambda/3$ and $\dot{H}=0$ with a
trivially satisfied  scalar equation. In this case, all
scalar-derivative terms disappear but its perturbation will keep the
nature of NDC+FK terms. This is a reason why we will choose the
simplistic background geometry of dS spacetime to study cosmological
implication of our action (\ref{enf}). It is interesting to note
that the Einstein+NDC+CC terms provides the dS form $a(t)=e^{Ht}$
with $H=(3\sqrt{\xi/M^2})^{-1}$ and $\phi(t)=e^{-t/\sqrt{\xi/M^2}}$
at $t=-\infty$~\cite{Sushkov:2009hk}.

 A solution of dS spacetime  to
Eqs.(\ref{einseq}) and (\ref{s-eq}) together with $\bar{T}^{\rm
NF}_{\mu\nu}=0$ can be easily found when one chooses a constant
scalar
\begin{equation} \label{dS-con}
\bar{R}=4\kappa \Lambda,~~\bar{\phi}={\rm const}.
\end{equation}
Here, the  curvature, Ricci, and Einstein  tensors can be written by
\begin{equation}
\bar{R}_{\mu\nu\rho\sigma}=H^2(\bar{g}_{\mu\rho}\bar{g}_{\nu\sigma}-\bar{g}_{\mu\sigma}\bar{g}_{\nu\rho}),~~\bar{R}_{\mu\nu}=3H^2\bar{g}_{\mu\nu},~~\bar{G}_{\mu\nu}=-3H^2\bar{g}_{\mu\nu}
\end{equation}
with  Hubble constant $H=\sqrt{\kappa \Lambda/3}$. Further, the
flat-slicing of dS spacetime can be realized  by  introducing
conformal time $\eta$ as
\begin{eqnarray} \label{dS}
\label{deds2} ds^2_{\rm dS}=\bar{g}_{\mu\nu}dx^\mu dx^\nu
=a(\eta)^2[-d\eta^2+\delta_{ij}dx^idx^j],
\end{eqnarray}
where  $a(\eta)$ is  conformal  scale factor expressed by
\begin{eqnarray}
a(\eta)=-\frac{1}{H\eta}.
\end{eqnarray}
 During dS inflation , the scale factor $a$ goes from small to a very
large value like $a_f / a_i\simeq 10^{30}$ which implies that the
conformal time $\eta=-1/aH(z=-k\eta)$ runs from $-\infty(\infty)$
[the infinite past] to $0^-(0)$ [the infinite future]. However, the
termination of inflation is not controlled by the cosmological
constant and thus, the inflation continues eternally. In other
words,  there is no mechanism to exit from inflation.  Even though
the dS inflation has  such a handicap in compared to slow-roll
inflation (quasi-dS inflation), we choose the dS inflation because
our model (\ref{enf}) seems unlikely to provide the slow-roll
inflation.

 The dS SO(1,4)-invariant distance between two spacetime points $x^\mu$ and $x'^{\mu}$ is defined  by
\begin{eqnarray}
\label{id2}
Z(x,x')=1-\frac{-(\eta-\eta')^2+|{\bf{x}}-{\bf{x}'}|^2}{4\eta\eta'}=1-\frac{(x-x')^2}{4\eta\eta'}
\end{eqnarray}
since $Z(x,x')$ has the ten symmetries which leave the metric of dS
spacetime invariant. Here $(x-x')^2$ is the Lorentz-invariant flat
spacetime distance.

\section{Scalar fourth-order propagator}

One begins  with general perturbed metric with 10 DOF
\begin{equation} \label{gpm}
ds^2=a(\eta)^2\Big[-(1+2\Psi)d\eta^2+2B_i d\eta
dx^{i}+(\delta_{ij}+\bar{h}_{ij})dx^idx^j\Big],
\end{equation}
where the SO(3)-decomposition is given by
\begin{eqnarray}
B_i=\partial_iB
+\Psi_i,~\bar{h}_{ij}=2\Phi\delta_{ij}+2\partial_{ij}E+\partial_i\bar{E}_j+\partial_j\bar{E}_i+h_{ij}
\end{eqnarray}
with  the transverse vectors $\partial_i\Psi^i=0$,
$\partial_i\bar{E}^i=0$, and transverse-traceless tensor
$\partial_ih^{ij}=h=0$. To investigate the cosmological perturbation
around the dS spacetime (\ref{deds2}), we might  choose the
Newtonian gauge as $B=E=0 $ and $\bar{E}_i=0$. Under this gauge, the
corresponding perturbed metric with 6 DOF and perturbed scalar can
be written as
\begin{eqnarray} \label{so3ds}
ds^2&=&a(\eta)^2\Big[-(1+2\Psi)d\eta^2+2\Psi_i d\eta
dx^{i}+\Big\{(1+2\Phi)\delta_{ij}+h_{ij}\Big\}dx^idx^j\Big],\\
\phi&=& \bar{\phi}+\varphi\label{phip}.
\end{eqnarray}
Now we linearize the Einstein equation (\ref{einseq}) around the dS
background to obtain  the cosmological perturbed equations. It is
known that the tensor perturbation  is decoupled from scalars. The
tensor  equation becomes a tensor form of a massless scalar equation
\begin{eqnarray}
\delta R_{\mu\nu}(h)-3H^2h_{\mu\nu}=0 \to
\bar{\nabla}^2h_{ij}=0.\label{heq}
\end{eqnarray}
We mention briefly how do two scalars $\Psi$ and $\Phi$, and a
vector $\Psi_i$ go on. The linearized Einstein equation requires
$\Psi=-\Phi$ which was  used to define the comoving curvature
perturbation in the slow-roll inflation and thus, they are not
physically propagating modes in dS spacetime.  During the dS
inflation, no coupling between $\{\Psi,\Phi\}$ and $\varphi$ occurs
since $\bar{\phi}=0$ implies $\delta T^{\rm NDC}_{\mu\nu}=\delta
T^{\rm FK}_{\mu\nu}=0$. Furthermore, the vector $\Psi_i$ is not  a
propagating mode in the ENF theory because it has no kinetic term.
Hence, we have the tensor $h_{ij}$  with 2 DOF propagating in dS
spacetime.

It would be better  to find the scalar power spectrum  by making
Fourier transform of propagator in dS spacetime. First of all, we
consider the  $\xi=2/3$ case. In this case, the perturbed scalar
equation takes the form
\begin{equation}\label{maseq}
\bar{\nabla}^4\varphi+\xi
\bar{G}^{\mu\nu}\bar{\nabla}_\mu\bar{\nabla}_\nu \varphi=0
 \to_{\xi=2/3}
 \bar{\nabla}^4\varphi-2H^2\bar{\nabla}^2\varphi\equiv
\Delta_4\varphi=0,
\end{equation}
where $\Delta_4$ is just the Weyl operator (conformally covariant
fourth-order operator) in dS
spacetime~\cite{Mazur:2001aa,Mottola:2010gp}.  This is the main
reason why we have introduced our action of $S_{\rm ENF}$
(\ref{enf}).

Actually, the non-degenerate fourth-order equation can be factorized
as
\begin{equation}\label{fseq}
\bar{\nabla}^2(\bar{\nabla}^2-2H^2)\varphi=0,
\end{equation}
which implies two second-order equations for mmc  and mcc scalars
\begin{eqnarray}
\bar{\nabla}^2\varphi^{({\rm mmc})}&=&0,\label{eeq}\\
(\bar{\nabla}^2-2H^2)\varphi^{({\rm mcc})}&=&0\label{meq},
\end{eqnarray}
where the solution to (\ref{fseq}) is given by
$\varphi=\varphi^{({\rm mmc})}+\varphi^{({\rm mcc})}$. For
simplicity, $(i)$ denotes two cases: ($i$=1) for mmc and ($i$=2) for
mcc.  We emphasize that a choice of $\xi=2/3$ leads to a mcc scalar.
Otherwise, one has a massive scalar propagating on dS spacetime.

Expanding $\varphi^{(i)}$ in terms of Fourier modes $\phi^{(i)}_{\bf
k}(\eta)$
\begin{eqnarray}\label{sfour}
\varphi^{(i)}(\eta,{\bf x})=\frac{1}{(2\pi)^{\frac{3}{2}}}\int
d^3{\bf k}~\phi^{(i)}_{\bf k}(\eta)e^{i{\bf k}\cdot{\bf x}},
\end{eqnarray}
Eqs.(\ref{eeq}) and (\ref{meq}) become
\begin{eqnarray}\label{s-eq2}
\Big(\frac{d^2}{d z^2}-\frac{2}{z}\frac{d}{d
z}+1\Big)\phi^{({1})}_{\bf
k}&=&0,\label{pmsoll}\\
\Big(\frac{d^2}{dz^2}-\frac{2}{z}\frac{d}{d
z}+1+\frac{2}{z^2}\Big)\phi^{(2)}_{\bf k}&=&0\label{pmmsoll}
\end{eqnarray}
with  $z=-\eta k$. Solutions to (\ref{pmsoll}) and (\ref{pmmsoll})
are given by
\begin{eqnarray}
\phi_{\bf k}^{(1)}&=&c_1(i+z)e^{iz},\label{pmsoll1}\\
\phi_{\bf k}^{(2)}&=&c_2ize^{iz},\label{pmmsoll1}
\end{eqnarray}
where $c_{1}$ and $c_2$ are constants to be determined. These will
be used to compute the power spectrum directly in the next section.

On the other hand,   the linearized equation with an external source
$J_{\varphi}$ takes the form
\begin{equation}
\Delta_4\varphi=-M^2J_{\varphi} \to
\varphi(x)=-\frac{M^2}{\Delta_4}J_{\varphi}\equiv
-D(Z(x,x'))J_{\varphi}(x'),
\end{equation}
where the propagator is given by the inverse of $\Delta_4$
as~\cite{Antoniadis:2011ib}
\begin{equation} \label{4thp}
D(Z(x,x'))=\frac{M^2}{2H^2}\Big[\frac{1}{-\bar{\nabla}^2}-\frac{1}{-\bar{\nabla}^2+2H^2}\Big]
=\frac{M^2}{2H^2}[G_{\rm mmc}(Z(x,x'))-G_{\rm mcc}(Z(x,x'))]
\end{equation}
with the dS-invariant distance $Z(x,x')$ (\ref{id2}). Here the
propagators of mmc scalar~\cite{Bros:2010wa} and mcc
 scalar~\cite{Higuchi:2009ew} in dS
spacetime are given by
\begin{equation}\label{prop}
G_{\rm
mmc}(Z(x,x'))=\frac{H^2}{(4\pi)^2}\Big[\frac{1}{1-Z}-2\ln(1-Z)+c_0\Big],~~G_{\rm
mcc}(Z(x,x'))=\frac{H^2}{(4\pi)^2}\frac{1}{1-Z},
\end{equation}
where the former is the dS invariant renormalized two-point function
(on the space of non-constant modes), while the latter is the
simplest scalar two-point function on dS spacetime. Substituting
(\ref{prop}) into (\ref{4thp}), the propagator takes the form
\begin{equation} \label{log-pro}
D(Z(x,x'))=\frac{M^2}{16\pi^2}\Big(-\ln[1-Z(x,x')]+\frac{c_0}{2}\Big)
\end{equation}
 which is a purely logarithm up to an additive constant
 $c_0$.

 The scalar power spectrum is defined by Fourier transforming
 the propagator at equal time $\eta=\eta'$ as
given by \begin{eqnarray}  {\cal P}&=&\frac{1}{(2\pi)^3}\int d^3{\bf
r} ~4\pi k^3 D(Z({\bf x},\eta;{\bf x}',\eta))e^{-i{\bf k}\cdot {\bf
r}},~~{\bf r}={\bf x}-{\bf x}'\label{cesa1}\\
&=&\frac{1}{(2\pi)^3}\frac{k^3M^2}{4\pi}\int d^3{\bf r}
\Big(-\ln\Big[\frac{r^2}{4\eta^2}\Big]+\frac{c_0}{2}\Big)e^{-i{\bf
k}\cdot {\bf
r}}\label{cesa2}\\
&=&-\frac{1}{(2\pi)^3}\frac{k^3M^2}{4\pi}\int d^3{\bf
r}\ln[r^2]e^{-i{\bf k}\cdot {\bf
r}}+\frac{k^3M^2}{4\pi}\Big(\ln[4\eta^2]+\frac{c_0}{2}\Big)\delta^3({\bf k})\label{cesa3}\\
&=&-\frac{M^2k^2}{8\pi^3}\int_0^{\infty} dr
\Big\{r\sin[kr]\ln[r^2]\Big\},\label{cesa4}
\end{eqnarray}
where the last (time-dependent) term in (\ref{cesa3}) disappeared,
thanks to \begin{equation} k^3\delta^3({\bf
k})=\frac{k\delta(k)\delta(\theta)\delta(\phi)}{\sin \theta}=0.
\end{equation}
We may use  Ces\`{a}ro-summation method  to compute a
logarithmically divergent integral
(\ref{cesa4})~\cite{Youssef:2012cx}. For this purpose,  we  note
that the integral of $\int_0^{\infty}f(x)dx$ is Ces\`{a}ro summable,
if
\begin{eqnarray}
(C,\alpha)=\lim_{\lambda\to\infty}\int_{0}^{\lambda}dx
\left(1-\frac{x}{\lambda}\right)^{\alpha}f(x)
\end{eqnarray}
exists and is finite for  integer   $\alpha\ge0$. Then, $(C,\beta)$
is also Ces\`{a}ro summable for any integer $\beta>\alpha$.

To investigate Ces\`{a}ro-summability of the integral explicitly, we
focus on  $f(r)=r\sin[kr]\ln[r^2]$ in (\ref{cesa4}). In this case,
$(C,\alpha)$ is
\begin{eqnarray}\label{ca}
(C,\alpha)=\lim_{\lambda\to\infty}\int_{0}^{\lambda} dr
\left(1-\frac{r}{\lambda}\right)^{\alpha}
\Big(r\sin[kr]\ln[r^2]\Big).
\end{eqnarray}
We note that  the integral $(C,0)$ takes the same form as
(\ref{cesa4}). After manipulations, we  have
\begin{eqnarray}\label{c0}
(C,0)=-\frac{1}{k^2}\lim_{\lambda\to\infty}\Big[\Big(k\lambda\cos[k\lambda]
-\sin[k\lambda]\Big)\ln[\lambda^2/2]+{\rm Si}[k\lambda]\Big],
\end{eqnarray}
where Si$[x]$ denotes the sine-integral function defined by
Si$[x]=\int_{0}^{x}(\sin[t]/t)dt$. The first two terms in (\ref{c0})
diverge  in the $\lambda\to\infty$ limit and thus, $(C,0)$ is not a
convergent integral. However, the last term in (\ref{c0}) is finite
as it is shown in
\begin{eqnarray}\label{sintt}
\lim_{\lambda\to\infty}\int_{0}^{k\lambda}\frac{\sin[t]}{t}dt=\frac{\pi}{2}.
\end{eqnarray}
Further, $(C,1)$ is also not convergent since it  becomes
\begin{eqnarray}\label{c1}
(C,1)=-\frac{1}{k^2}\lim_{\lambda\to\infty}\Big(\sin[k\lambda]\ln[\lambda^2/2]
+{\rm Si}[k\lambda]\Big),
\end{eqnarray}
where the first term  diverges  in the limit of $\lambda\to\infty$.
On the other hand, for $\alpha=2$, the corresponding integral
(\ref{ca}) has a finite value
\begin{eqnarray}\label{c2}
(C,2)=-\frac{2}{k^2}\lim_{\lambda\to\infty}\Big({\rm
Si}[k\lambda]\Big)=-\frac{\pi}{k^2},
\end{eqnarray}
where we used (\ref{sintt}).  Also, we have checked that in the
limit of $\lambda\to\infty$,
\begin{eqnarray}
(C,2)=(C,3)=(C,4)=(C,5)=\cdots.
\end{eqnarray}
This implies that $(C,\beta)$ is also Ces\`{a}ro summable for any
integer $\beta>\alpha=2$.

Considering (\ref{cesa4}) together with (\ref{c2}), the scalar
spectrum takes the form
 \begin{equation}
{\cal P}=\frac{M^2}{8\pi^2}.
\end{equation}
For $M^2=2H^2$, it leads to an exactly  scale-invariant spectrum
 \begin{equation} \label{sc-inv}
{\cal P}=\Big(\frac{H}{2\pi}\Big)^2.
\end{equation}
This can be easily  checked by noting that for $M^2=2H^2$ and
$\xi=2/3$, $S_{\rm NF}$ in (\ref{enf}) becomes the Lee-Wick scalar
model in dS spacetime~\cite{Myung:2014mla}.

If one uses (\ref{4thp}) instead of (\ref{log-pro}) with $M^2=2H^2$,
after Fourier transforming it, its spectrum is computed to be
\begin{equation} \label{f-ps}
{\cal P}_{\rm f}={\cal P}_{\rm mmc}-{\cal P}_{\rm
mcc}=\Big(\frac{H}{2\pi}\Big)^2\Big[1+(k\eta)^2-(k\eta)^2\Big]
=\Big(\frac{H}{2\pi}\Big)^2,
\end{equation}
which is a unitary scale-invariant spectrum.

Finally, we would like to mention the unfixed $\xi$ case  whose
fourth-order equation is given by
\begin{equation}\label{maseqg}
\bar{\nabla}^4\varphi+\xi
\bar{G}^{\mu\nu}\bar{\nabla}_\mu\bar{\nabla}_\nu \varphi=0
\to\bar{\nabla}^4\varphi-m_{\xi}^2\bar{\nabla}^2\varphi\equiv
\Delta^\xi_4\varphi=0
\end{equation}
with the mass squared  \begin{equation} m^2_\xi=3\xi H^2.
\label{mxi}\end{equation}
 Its
propagator takes the form
\begin{equation} \label{4thpg}
D[Z(x,x'),m^2_\xi]=\frac{M^2}{2H^2}\Big[\frac{1}{-\bar{\nabla}^2}-\frac{1}{-\bar{\nabla}^2+m^2_\xi}\Big]
=\frac{M^2}{2H^2}\Big[G_{\rm mmc}(Z(x,x'))-G[Z(x,x'),m^2_\xi]\Big].
\end{equation}
 In general, a
massive scalar propagator $G[Z;m^2_\xi]$ must depend on the
SO(1,4)-invariant distance $Z(x,x')$ (\ref{id2}) which has ten
symmetries in dS spacetime. Also, it should satisfy the scalar wave
equation for $x\not=x'~(Z\not=1)$
\begin{equation} \label{green-eq}
\Big[Z(1-Z)\frac{d^2}{dZ^2}+2(1-2Z)\frac{d}{dZ}-\frac{m_\xi^2}{H^2}\Big]G[Z(x,x');m_\xi^2]=0
\end{equation}
whose solution is given by the hypergeometric function
\begin{equation}  \label{green-fn}
G[Z;m^2_\xi]=\frac{H^2}{16\pi^2}\Gamma(\frac{3}{2}+\nu_\xi)\Gamma(\frac{3}{2}-\nu_\xi)~
{}_2F_1\Big[\frac{3}{2}+\nu_\xi,\frac{3}{2}-\nu_\xi,2;Z\Big],~\nu_\xi=\sqrt{\frac{9}{4}-\frac{m_\xi^2}{H^2}}.
\end{equation}
At $x=x'$, (\ref{green-fn}) is correctly normalized for
(\ref{green-eq}) to give $\delta^4(x,x')$ with unit weight. We
observe that $\nu_{\xi}\ge 0 \to 0<\xi \le \frac{3}{4}$. The
negative $\xi$ is not allowed because it gives us tachyon.  Here we
note that $G[Z,m^2_\xi=0]\to G_{\rm mmc}(Z)$ and
$G[Z,m^2_{\xi=2/3}=2H^2]\to G_{\rm mcc}(Z)$. However, it is a
formidable task to obtain its scalar power spectrum by Fourier
transforming (\ref{4thpg}) at $\eta=\eta'$ because it is difficult
to Fourier transform $G[Z({\bf x},\eta;{\bf x}',\eta),m^2_\xi]$ with
$Z=1-\frac{|{\bf x}-{\bf x}'|^2}{4\eta^2}$. In the next section, we
could compute the scalar power spectrum for arbitrary $\xi\le 3/4$
directly.
\section{Scalar spectrum}

In order to compute  scalar power spectrum directly, we have to
obtain the second-order bilinear action. Making use of the
Ostrogradski's formalism, one may rewrite the fourth-order bilinear
action $\delta S_{\rm NF}$ obtained by bilinearizing (\ref{enf}) as
the second-order bilinear action
\begin{eqnarray}
&&\hspace*{-2.3em}\delta S^{(2)}_{\rm NF}=\frac{1}{2M^2}\int
d^4x\Big[a^2
H^2\Big((3\xi-4)\alpha^2-3\xi\partial_i\varphi\partial^i\varphi\Big)
-\Big((\alpha')^2-2\partial_i\alpha\partial^i\alpha
+\partial^2\varphi\partial^2\varphi\nonumber\\
&&\hspace*{8em}+4aH\alpha\alpha'-4aH\alpha\partial^2\varphi\Big)
+2M^2\lambda(\alpha-\varphi')\Big]\label{sndc},
\end{eqnarray}
where $\alpha\equiv \varphi'$ is a new variable to lower
fourth-order derivative down  and $\lambda$ is a Lagrange
multiplier. In (\ref{sndc}), the prime ($'$) denotes the
differentiation with respect to $\eta$.

From the action (\ref{sndc}), the conjugate momenta are given by
\begin{eqnarray} \label{conju}
\pi_{\varphi}=-\lambda,
~~~\pi_{\alpha}=-\frac{1}{M^2}(\alpha'+2aH\alpha).
\end{eqnarray}
Varying the action (\ref{sndc}) with respect to $\lambda$ and
$\alpha$ leads to equations
\begin{equation} \label{abmon}
\alpha=\varphi',~~
\lambda=-\frac{1}{M^2}\Big(\alpha''+a^2H^2(3\xi-2)\varphi'-2\partial^2\alpha+2aH\partial^2\varphi\Big).
\end{equation}
Plugging (\ref{abmon}) into (\ref{conju}), the conjugate momenta are
given by
\begin{eqnarray}\label{conjm}
\pi_{\varphi}=\frac{1}{M^2}\Big(\varphi'''+a^2H^2(3\xi-2)\varphi'-2\partial^2\varphi'+2aH\partial^2\varphi\Big),
~~\pi_{\alpha}=-\frac{1}{M^2}(\varphi''+2aH\varphi').
\end{eqnarray}
The canonical quantization is accomplished by imposing two
commutation relations
\begin{eqnarray}\label{comr}
[\hat{\varphi}(\eta,{\bf x}),~\hat{\pi}_{\varphi}(\eta,{\bf
x}^{\prime})]=i\delta({\bf x}-{\bf
x}^{\prime}),~~~[\hat{\alpha}(\eta,{\bf
x}),~\hat{\pi}_{\alpha}(\eta,{\bf x}^{\prime})]=i\delta({\bf x}-{\bf
x}^{\prime}).
\end{eqnarray}
The field  operator $\hat{\varphi}$ can be expanded in Fourier modes
as
\begin{eqnarray}\label{phiex}
\hat{\varphi}(\eta,{\bf x})=\frac{1}{(2\pi)^{\frac{3}{2}}}\int
d^3{\bf k}\Big[\Big(\hat{a}_{\bf k}\phi_{\bf
k}^{(1)}(\eta)+\hat{b}_{\bf k}\phi_{\bf k}^{(2)}(\eta)\Big)e^{i{\bf
k}\cdot{\bf x}}+~{\rm h.c.}\Big],
\end{eqnarray}
where $\phi_{\bf k}^{(1)}$ and $\phi_{\bf k}^{(2)}$ were given by
(\ref{pmsoll1}) and (\ref{pmmsoll1}). When one substitutes
(\ref{phiex}) into the operator of $\hat{\pi}_{\varphi}$,
$\hat{\alpha}(\equiv\hat{\varphi}')$, and $\hat{\pi}_{\alpha}$, one
obtains the corresponding expressions. Plugging these all into
(\ref{comr}), two commutation relations take the forms
\begin{eqnarray}
&&\hspace*{-2em}[\hat{a}_{\bf k},~\hat{a}_{\bf
k^{\prime}}^{\dag}]=\delta({\bf k}-{\bf
k}^{\prime}),~~~~~~[\hat{b}_{\bf k},~\hat{b}_{\bf
k^{\prime}}^{\dag}]=-\delta({\bf k}-{\bf
k}^{\prime}).\label{comrell}
\end{eqnarray}
It is noted that two mode operators ($\hat{a}_{\bf k},\hat{b}_{\bf
k}$) are necessary to take into account of fourth-order theory
quantum mechanically  as the Pais-Uhlenbeck fourth-order oscillator
has been  shown in~\cite{Mannheim:2004qz}. We remind the reader that
the unusual commutator for ($\hat{b}_{\bf k},\hat{b}_{\bf
k^{\prime}}^{\dag}$) reflects that the FK term contains the ghost
state scalar~\cite{Chen:2013aha}. In addition, two Wronskian
conditions are found to be
\begin{eqnarray}
&&\hspace*{-2em}\Big[\phi_{\bf k}^{(1)}\Big\{\Big(\phi_{\bf
k}^{*(1)}(\eta)\Big)'''+\Big((3\xi-2)a^2H^2+2k^2\Big)\Big(\phi_{\bf
k}^{*(1)}(\eta)\Big)'-2aHk^2\phi_{\bf
k}^{*(1)}(\eta)\Big\}\nonumber\\
&&\hspace*{-2em}-\phi_{\bf k}^{(2)}\Big\{\Big(\phi_{\bf
k}^{*(2)}(\eta)\Big)'''+\Big((3\xi-2)a^2H^2+2k^2\Big)\Big(\phi_{\bf
k}^{*(1)}(\eta)\Big)'-2aHk^2\phi_{\bf
k}^{*(2)}(\eta)\Big\}\Big]-~c.c.=iM^2,\nonumber\\
&&\hspace*{-2em}\Big[\Big(\phi_{\bf
k}^{(1)}\Big)'\Big\{\Big(\phi_{\bf
k}^{*(1)}(\eta)\Big)''+2aH\Big(\phi_{\bf
k}^{*(1)}(\eta)\Big)'\Big\}\nonumber\\
&&\hspace*{5em}-\Big(\phi_{\bf k}^{(2)}\Big)'\Big\{\Big(\phi_{\bf
k}^{*(2)}(\eta)\Big)''+2aH\Big(\phi_{\bf
k}^{*(2)}(\eta)\Big)'\Big\}\Big]-c.c.=-iM^2,\label{wconb}
\end{eqnarray}
which will be used  to fix the coefficients of solutions $\phi_{\bf
k}^{(1)}$ and $\phi_{\bf k}^{(2)}$ to two second-order equations
\begin{eqnarray}
\bar{\nabla}^2\phi_{\bf k}^{(1)}&=&0,\\
 (\bar{\nabla}^2-m_{\xi}^2)\phi_{\bf
k}^{(2)}&=&0,
\end{eqnarray}
where $m_{\xi}^2$ is given by $(\ref{mxi})$. Explicitly, the
solutions are determined to be
\begin{equation}
\phi_{\bf
k}^{(1)}=\frac{M}{m_\xi}\frac{H}{\sqrt{2k^3}}(i+z)e^{iz},~~
\phi_{\bf
k}^{(2)}=\frac{M}{m_{\xi}}\frac{H}{\sqrt{2k^3}}z^{\frac{3}{2}}H_{\nu_{\xi}}(z)\label{p2sol2},
\end{equation}
where $H_{\nu_\xi}^{(1)}$ is the first-kind Hankel function  and its
index $\nu_\xi$ is given  by (\ref{green-fn}).

 On the other hand,
the power spectrum of the scalar is defined by~\cite{Baumann:2009ds}
\begin{eqnarray}\label{pow}
\langle0|\hat{\varphi}(\eta,{\bf x})\hat{\varphi}(\eta,{\bf
x^{\prime}})|0\rangle=\int d^3{\bf k}\frac{{\cal
P}_{\varphi}(k,\eta)}{4\pi k^3}e^{i{\bf k}\cdot({\bf x}-{\bf
x^{\prime}})}.
\end{eqnarray}
Considering the Bunch-Davies vacuum state imposed by  $\hat{a}_{\bf
k}|0\rangle=0 $ and $\hat{b}_{\bf k}|0\rangle=0$, (\ref{pow}) is
computed  as
\begin{eqnarray}
{\cal P}_{\rm
\varphi}(k,\eta)&=&\frac{k^3}{2\pi^2}\left(\Big|\phi_{\bf
k}^{(1)}\Big|^2-\Big|\phi_{\bf k}^{(2)}\Big|^2\right)\label{powf}\\
&=&\frac{M^2}{12\xi\pi^2}\Big[(1+z^2)-\frac{\pi}{2}z^3|H_{\nu_\xi}^{(1)}(z)|^2\Big]\label{powff},
\end{eqnarray}
where we have used (\ref{mxi}). Importantly, the minus sign ($-$) in
(\ref{powf}) appears because the unusual commutation relation
($\hat{b}_{\bf k},\hat{b}_{\bf k^{\prime}}^{\dag}$) for ghost state
was used. We expect to derive the power spectrum (\ref{powff}) by
Fourier transforming (\ref{4thpg}) at $\eta=\eta'$.

For $\xi=2/3(\nu_\xi=1/2)$,  it is shown that considering
\begin{eqnarray}
H_{1/2}^{(1)}(z)=\sqrt{\frac{2}{\pi}}z^{-1/2}e^{i (z-\pi/2)},
\end{eqnarray}
the power spectrum (\ref{powff}) is found to be
\begin{eqnarray} \label{scalar-pow}
{\cal P}_{\rm
\varphi}~=~\frac{M^2}{8\pi^2}\Big[1+\frac{k^2}{(aH)^2}-\frac{k^2}{(aH)^2}\Big]~=~\frac{M^2}{8\pi^2}.
\end{eqnarray}
For $M^2=2H^2$, it leads to  the HZ scale-invariant power spectrum
\begin{eqnarray}
{\cal P}_{\rm \varphi}^{M^2=2H^2}~=~\left(\frac{H}{2\pi}\right)^2,
\end{eqnarray}
which is just the same form as in (\ref{sc-inv}).  We emphasize that
the minus sign in (\ref{scalar-pow}) is essential to derive a
scale-invariant spectrum from the scale-variant spectrum
(\ref{powff}).

Finally, we would like to mention that for $\xi\not=2/3$, its power
spectrum (\ref{powff}) becomes
\begin{eqnarray} \label{powffs}
{\cal P}_{\rm \varphi}(k,\eta)
=\frac{M^2}{12\xi\pi^2}\Big[1+\Big(\frac{k}{aH}\Big)^2-\frac{\pi}{2}\Big(\frac{k}{aH}\Big)^3|H_{\nu_\xi}^{(1)}\Big(\frac{k}{aH}\Big)|^2\Big]
\label{powff1},
\end{eqnarray}
which is obviously a  scale-variant spectrum for dS inflation. In
the superhorizon limit of $z\to 0(k\ll aH)$, (\ref{powff1}) takes
the form
\begin{eqnarray}
{\cal P}_{\rm \varphi}(k,\eta)|_{z\to 0}
=\frac{M^2}{12\xi\pi^2}\Big[1-\Big(\frac{\Gamma(\nu_\xi)}{\Gamma(3/2)}\Big)^2\Big(\frac{k}{2aH}\Big)^{3-2\nu_\xi}\Big]
\label{powff2}
\end{eqnarray}
which is not scale-invariant.

\section{Discussions}

We have obtained an exactly scale-invariant spectrum of scalar
perturbation generated during de Sitter inflation  from the gravity
model of the nonminimal derivative coupling with fourth-order term.
This is the case of $\xi=2/3$ where there was cancellation between
healthy massless scalar and unhealthy (ghost) scalar. For
$\xi\not=2/3$, we have obtained the scale-variant spectrum
(\ref{powff1}) where there was no cancellation between healthy
massless scalar and unhealthy  massive scalar.

 The nonminimal derivative coupling term
generates a healthy  second-order term of
$-2H^2\bar{\nabla}^2\varphi$ for $\xi=2/3$, while the fourth-order
term provides an unhealthy  fourth-order derivative term. This
combination provided the linearized scalar equation of
$\Delta_4\varphi=0$ expressed in term of the Weyl fourth-order
operator $\Delta_4$ in dS spacetime. In this sense, our model
$S_{\rm NF}$ of NDC with FK term is more promising  than the LW
scalar model (\ref{lw}) where the NDC term is replaced by CC term of
$-\bar{\nabla}^2\varphi$.

Now we explain  a consistency of  our  perturbation theory in dS
spacetime. The dS-invariant correlation function (\ref{green-fn})
was constructed to possess full conformal invariance.   We note that
our propagator (\ref{log-pro}) is a logarithmic correlator with the
zero conformal weight. This means that our model  (\ref{enf}) with
$\xi=2/3$ may be considered as a minimal model which is allowed by
unitarity of quantum field theory in dS spacetime. Our fluctuations
which are similar to  the anomaly scalar
fluctuations~\cite{Mottola:2010gp,Antoniadis:2011ib} can give rise
to the HZ scale-invariant power spectrum (\ref{sc-inv}) in a fully
SO(3,1) conformally invariant way on the dS horizon. This implies
that  that (\ref{log-pro}) can  give  the unitary scale-invariant
power spectrum (\ref{f-ps})  in dS spacetime. However, our model
with $\xi\not=2/3$ has led to the scale-variant spectrum
(\ref{powffs}) which violates the unitarity.

The HZ scale-invariant
 spectrum  was obtained from Fourier transforming the fourth-order propagator (\ref{log-pro})  in
de Sitter spacetime. Taming a logarithmic IR divergence by making
use of the Ces\`{a}ro summability technique, we arrived at the power
spectrum of $(H/2\pi)^2$.   Importantly, this HZ spectrum  was also
recovered by computing the power spectrum in its momentum space
directly. In obtaining  the power spectrum,  we have used the
Ostrogradski's formalism and quantization scheme of Pais-Uhlenbeck
fourth-order oscillator. Hence, it is argued that our model
(\ref{enf}) can provide  a consistent perturbation theory on the dS
background. Here, we remark that in the case of $M^2=1$, one has a
scalar with zero mass dimension in (\ref{enf}). This corresponds to
the case (7.3) in~\cite{Antoniadis:2011ib}.   In this case, one has
a constant power spectrum of ${\cal P}=1/(8\pi^2)$ which seems to be
trivial.

Finally, we would like to mention the ghost issue because the FK
term generates ghost problem.  It is well known that NDC term does
not contain a ghost due to the use of the Einstein tensor as a
coupling function, while other NDC terms could contain
ghosts~\cite{Sushkov:2009hk}. We discuss this issue by separating
the dS background from the perturbation around dS background. This
suggests strongly that one has to distinguish  higher-time
derivative terms (ghost) from ghost state (negative-norm state). As
was observed from (\ref{Heq}), it is evident that $\rho_{\rm NDC}$
does not contain any dangerous higher-time derivative terms (ghost),
while $\rho_{\rm FK}$ involves fourth-order time derivatives
(ghost). In the case of choosing dS background ($H$=const,
$V=\Lambda$, $\phi$=const), all time-derivatives of $H$ and $\phi$
disappear. This implies that we never worry about the ghost issue on
the background evolution. In the case of slow-roll inflation
(quasi-dS), however,  the ghost problem arises due to the presence
of $\rho_{\rm FK}$. Going back to the perturbed scalar equation
(\ref{maseqg}), its fourth-order propagator (\ref{4thpg}) shows the
ghost state (negative sign in front of $G[Z(x,x'),m^2_\xi]$).
However, for $\xi=2/3$, its propagator reduces to the logarithmic
form (\ref{log-pro}) in dS spacetime which shows no ghost states
even though minus sign is present.  By Fourier transforming it at
equal-time leads to the scale-invariant spectrum (\ref{sc-inv}) for
$M^2=2H^2$ which is free from ghost state (negative-norm state).
This was confirmed by computing the power spectrum directly:
(\ref{powffs})$\to$(\ref{scalar-pow}). Even though the FK term gives
rise to  ghost state apparently, a special fourth-order propagator
(\ref{scalar-pow})  could be represented by a difference of green
functions between mmc (massless minimally coupled) scalar and mcc
(massless conformally coupled) scalar propagating in dS background.
The former corresponds to the dS-invariant renormalized two-point
function, whereas the latter represents the simplest scalar
two-point function in dS spacetimes. This is closely related to the
fact that $\Delta_4$ in (\ref{maseq}) becomes the Weyl operator
(conformally covariant fourth-order operator in dS Spacetime).  This
difference indicates the power spectrum (\ref{f-ps}) which is
clearly ghost-free (positive-norm state) and scale-invariant.
However, it seems unlikely to obtain the ghost-free propagator from
the same operator of $\partial^2(\partial^2-m^2)$ in Minkowski
spacetime \cite{Barth:1983hb}.

\vspace{0.35cm}

{\bf Acknowledgement}

\vspace{0.25cm}
 This work was supported by the National
Research Foundation of Korea (NRF) grant funded by the Korea
government (MEST) (No.2012-R1A1A2A10040499).

\end{document}